\documentclass[amssymb,prb,twocolumn,showpacs]{revtex4}
\usepackage{epsfig}
\usepackage{dcolumn}
\begin{document}
\title{\bf\ Radiation-induced resistance oscillations
 in 2D electron systems with strong Rashba coupling}

 \author{Jes\'us I\~narrea$^{1,2}$ }
\affiliation{$^1$Escuela Polit\'ecnica
Superior,Universidad Carlos III,Leganes,Madrid,Spain \\and
$^2$Unidad Asociada al Instituto de Ciencia de Materiales, CSIC
Cantoblanco,Madrid,28049,Spain.}
\date{\today}
\begin{abstract}
We present a theoretical study on the effect of radiation
on the mangetoresistance of two-dimensional electron systems with
strong Rashba spint-orbit coupling. We want to study the interplay between
two well-known effects in these electron systems: the radiation-induced
resistance oscillations and the typical beating pattern of systems with intense Rashba interaction.
We analytically derive an exact
solution for the electron wave function corresponding to a total Hamiltonian with
 Rashba and radiation terms. We consider
 a perturbation treatment for elastic scattering due
to charged impurities to finally obtain the
magnetoresistance of the system. Without radiation we recover a beating
pattern in the amplitude of the Shubnikov de Hass oscillations: a set of
nodes and antinodes in the magnetoresistance. In the presence of radiation
this beating pattern is strongly modified following the profile of radiation-induced
magnetoresistance oscillations. We study
their dependence on  intensity and frequency of
radiation, including the teraherzt regime. The obtained results could be of
interest for magnetotransport of nonideal Dirac fermions in 3D topological insulators
subjected to radiation.
\end{abstract}
\maketitle
\section{Introduction}
Radiation-induced resistance oscillations (RIRO) and zero resistance states (ZRS)\cite{mani1,zudov1}
are remarkable phenomena in condensed matter physics that reveal a
novel scenario in radiation-matter coupling. Those effects rise up
when a high-mobility, typically above $10^{6}$ $cm^{2}/Vs$, two dimensional electron system  under a
moderate magnetic field ($B$) is illuminated with microwave (MW) or
terahertz (TH) radiation. The mangetoresistance ($R_{xx}$) of such systems (2DES)
 shows oscillations with peaks and valleys at a certain
radiation power. When increasing  power, the $R_{xx}$ oscillations
increase in turn and at high enough intensity the valleys turn into ZRS.
The radiation-induced resistance oscillations
show characteristic traits such as periodicity in the inverse of $B$\cite{mani1,zudov1},
a 1/4 cycle shift in the oscillations minima\cite{mani2}, sensitivity to temperature
\cite{mani3,ina1} and radiation power\cite{mani4}. For the latter
case, a sublinear law is obtained for the dependence of RIRO on
the radiation power, $R_{xx} \propto P^{\alpha}$, where $P$ is
the radiation power and, interestingly,  the exponent is around $0.5$.
This clearly indicates a squared root dependence.
 \begin{figure}
\centering\epsfxsize=3.5in \epsfysize=6.0in
\epsffile{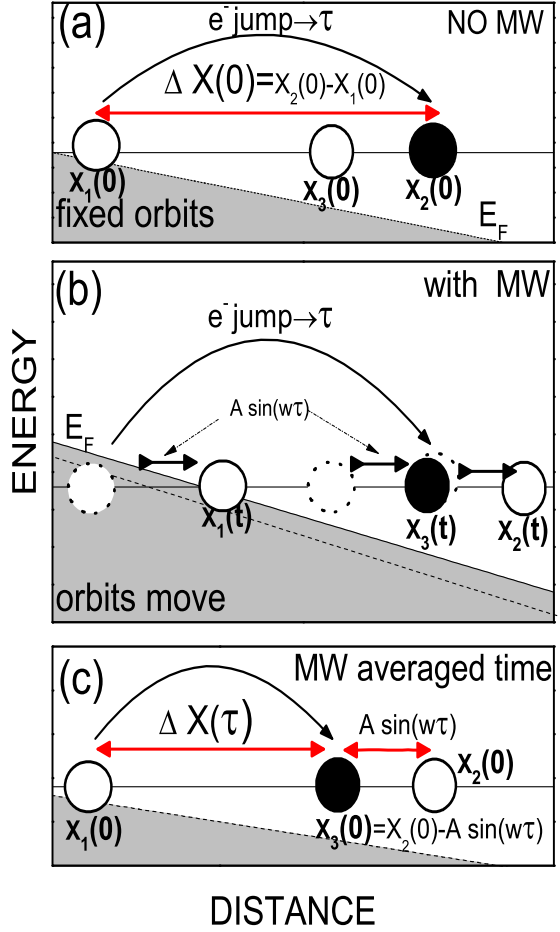}
\caption{ Schematic figure describing elastic scattering between Landau states or orbits.
a) Dark scenario. The scattering jump takes place between fixed orbits.  $\tau$ is the time
it takes the electron to complete the jump and reach the final orbit.
b) Under radiation the scattering jump is between oscillating orbits according
to $A \sin wt$. c) Scattering under radiation for an averaged time scenario, i.e.,
steady state. The stationary advanced distance turns out to be smaller than the dark one.
This corresponds to a valley in the radiation-induced magnetoresistance oscillations.}
\end{figure}

A great number of experiments and theoretical models
have been presented to date to try to explain such striking
effects. From a theoretical standpoint, we can cite for instance
the displacement model\cite{durst} based on radiation-assisted
inter Landau level scattering, the inelastic model based on the
effect of radiation on the nonequilibrium electron distribution function\cite{vavilov}.
Being these two models the most cited to date,
other models are
more successful explaining the basic features of RIRO, such as  the one by Lei et al\cite{lei},  or
the radiation-driven electron orbit model\cite{ina2,ina3,ina4,ina5,ina7}.
We have to admit that  to date there is no universally accepted theoretical approach among
the people devoted to this field. On the other hand, some experimental advances have been made ruling out
some of the previous existing theories. For instance a very recent experiment by
T.Herrmann\cite{herr} concludes that RIRO are mainly dependent on bulk effects
in 2DES. Then this experiment  rules out theories based on the effect of radiation on
edge states\cite{chepe}
 or others based on ponderomotive forces\cite{mija} excited by contact illumination, as main responsibles of RIRO.
  Another example, the experiments by Mani et al.,\cite{mani5,mani6,ye,ina6}
on power dependence  rule out the theories that claimed a linear dependence of oscillations on $P$
\cite{vavilov}.
 \begin{figure}
\centering\epsfxsize=3.8in \epsfysize=3.5in
\epsffile{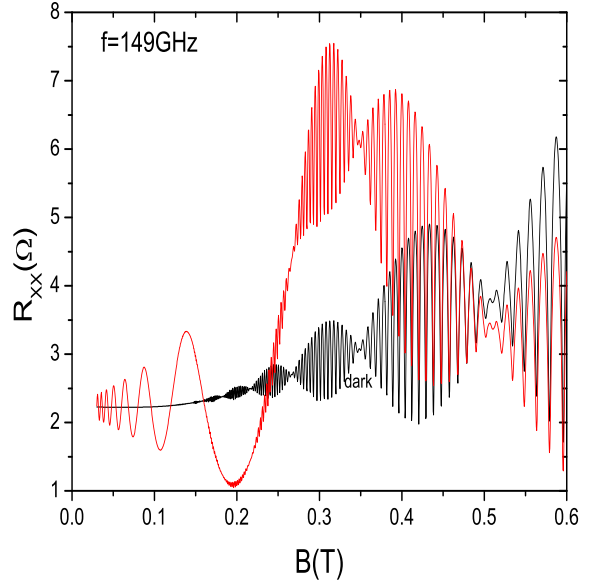}
\caption{ Calculated magnetoresistance, $R_{xx}$, vs magnetic field, $B$, for dark and radiation
of frequency $f= 149$ GHz, in a high-mobility 2DES with strong RSOI of Rashba parameter $\alpha=0.6\times 10^{-11}eV\cdot m$.
For the dark curve we obtain, as expected,  a beating pattern profile made up of
a system of nodes and antinodes. The radiation curve exhibits  similar beating
pattern but modulated by the rise
of the system of peaks and valleys of RIRO. (T=1K).}
\end{figure}
The current or future theoretical models dealing with RIRO or ZRS have to confront with
the available experimental results to prove how good and accurate they are.
Another approach to prove theories is to check out if they are able to predict results on novel scenarios where
there are no experiments carried out yet. For instance RIRO and ZRS obtained on different
semiconductor platforms other than GaAs, the most extensively platform used
in this kind of experiments. The main reason for using GaAs is that
this platform offers the highest mobility\cite{manfra} to date, ($\sim3.0\times 10^{7}$ $cm^{2}/Vs$), among different
semiconductor heterostructures.

In this article we present a theoretical study on the effect of radiation on
the magnetotransport in samples with strong Rashba\cite{vasco,rashba1,rashba2} spin-orbit interaction (RSOI),
such as InAs. The interest by heterostructures with InAs is increasing very fast in the last years, on the one hand for
their technological impact being part of new electronic devices\cite{pearton}. On the other hand
from basic research standpoint in fields such as spintronics transistors and the realisation of
Majorana fermions\cite{sarma1}. Usually, the electron mobility in InAs has been always
below $1.0\times 10^{6}$ $cm^{2}/Vs$ and RIRO can  hardly be seen.
Yet, there have been published very recently experimental results demonstrating that improving MBE growth techniques in
quantum wells of InAs  electron mobilities can be dramatically increased\cite{ihn}.
They claimed a mobility close to $3.0\times 10^{6}$ $cm^{2}/Vs$\cite{ihn}.
Therefore, samples of InAs, with strong RSOI,  can now become  reasonable candidates to
observe RIRO. Then, we could study  the interplay of Rashba interaction and radiation in
these kind of systems. We could also predict that with samples with even higher mobilities and
at high enough radiation intensity, 2DES systems with RSOI can
give rise to ZRS.

Thus,  we start off based on the previous  theory of {\it the radiation-driven electron
orbit}\cite{ina2,ina3,ina4,ina5,ina7}. This theory stems from the displacement model\cite{durst}
and shares with it  the interplay between charged impurity scattering and radiation to be at
the heart of RIRO. As a reminder, the displacement model
 includes, in a perturbative way, electron promotion with the corresponding photons absorption to higher Landau states and subsequent
electron advances and recoils via scattering to explain $R_{xx}$ peaks and valleys. Yet, the radiation-driven electron orbit theory
does not consider any radiation-assisted electron promotion. As a further evolution of the displacement model,
 our theory proceeds in an alternative approach starting from the
exact solution of the time-dependent Schrodinger equation for an electron under magnetic field and
radiation. The obtained exact wave function represents a Landau state where the guiding center
is harmonically driven back and forth by radiation at the same frequency. Interestingly, the Landau states guiding
center follow  a classical trajectory given
by the solution of the driven classical oscillator. According to this theory,
the interaction of the driven Landau states with charged impurities ends up giving rise to shorter
and longer  average advanced distance by the scattered electrons. These distance are reflected on irradiated
$R_{xx}$ as valleys and peaks respectively.  An interesting trait of this model is that it encompasses quantum and classical
concepts unlike the inelastic and the displacement models that are fully quantum.
At this point we can highlight a recent theoretical contribution
by Beltukov\cite{beltukov} et al. that is fully classical. It should not be surprising
to take into account and include totally or partially classical approaches explaining RIRO considering
the very low magnetic fields where this effect rise up.

 We have added to the same total Hamiltonian of the radiation-driven
electron orbit theory the Rashba interaction,
solving exactly the corresponding time-dependent Schrodinger equation. In other words, we develop
a semi-classical model based on the exact solution
of the electronic wave function in the presence of a static $B$ with
important Rashba coupling
interacting with radiation and a perturbation treatment for elastic scattering
from  charged impurities. We consider this type of scattering as the most
probable at the low temperatures normally used in experiments. Following the previous model and applying a Boltzmann transport model we are able to obtain an expression for
$R_{xx}$ with radiation and RSOI.
In the simulations we obtain, first without radiation, the well-known beating pattern with the system of nodes and
antinodes of the Rashba magnetoresistance\cite{vasilo1,vasilo2,nitta,shoja,heida,fete,rup,luo,luo2,das}. Then, we switched on light obtaining  $R_{xx}$ that shows a strong deformation
of the previous beating pattern where the nodes and antinodes follow the peaks and valleys of RIRO.
We study the dependence on power and frequency including
the terahertz regime. 2DES with RSOI share similar Hamiltonian with
3D topological insulators, then we consider that the
results that we present in this article could be of application
in the field of 3D topological insulators under the influence of radiation.

\section{Theoretical model}
 We consider a 2DES in the $x-y$ plane with strong Rashba coupling
subjected to a static and
perpendicular $B$ and a  DC electric field parallel to the $x$ direction.
Using Landau gauge for the potential vector, ${\bf A}=(0,Bx,0)$, the  hamiltonian of such a system, $H_{0}$ reads:
$H_{0}=(H_{B}+H_{SO})$
where the different components of  $H_{0}$ are:
\begin{eqnarray}
 H_{B}&=&\left[\frac{p_{x}^{2}}{2m^{*}}+\frac{1}{2}m_{*}w_{c}^2\left(x-X_{0} \right)^{2}\right]\sigma_{0}+\frac{1}{2}g\mu_{B}\sigma_{z}B+\nonumber\\
      & &\left[-eE_{dc}X_{0}+\frac{1}{2}m^{*}\frac{E_{dc}^{2}}{B^{2}}\right]\sigma_{0}
  \\
H_{SO}&=&-\frac{\alpha}{\hbar}\left[\sigma_{y}p_{x}-\sigma_{x}eB\left(x-X_{0}\right)\right]
\end{eqnarray}
 $X_{0}$ is the center of the orbit for the electron spiral motion:
$X_{0}=-\left(\frac{\hbar k_{y}}{eB}- \frac{eE_{dc}}{m^{*}w_{c}^{2}}\right)$,
 $E_{dc}$ is the DC
electric field parallel to the $x$ direction, $\sigma_{0}$ stands for
the unit matrix, $\overrightarrow{\sigma}=(\sigma_{x},\sigma_{y},\sigma_{z})$ are the
Pauli spin matrices, $g$ the Zeeman factor, $\mu_{B}$ the Bhor magneton,
$\alpha$ the Rashba spin-orbit coupling parameter and $w_{c}$ the cyclotron frequency.
The Schrodinger equation corresponding to the Hamiltonian  $H_{0}$ can be exactly solved and the resulting states are labeled by
the quantum number $N$. For $N=0$ there is only one level of energy
given by $E_{0}=(\hbar w_{c}-g\mu_{B} B)/2$. For $N\geq 1$ we
obtain two branches of levels labelled by
+ and - and with energies:
\begin{equation}
E_{N\pm}=\hbar w_{c}N\pm\frac{1}{2} \sqrt{(\hbar w_{c}-g\mu_{B} B)^{2}+\frac{8\alpha^{2}}{R^{2}}N}
\end{equation}
where $R$ is the magnetic length, $R=\sqrt{\frac{\hbar}{eB}}$.
The corresponding wave function for the + branch is,
\begin{equation}
\psi_{N+}=\frac{1}{\sqrt{L_{y}}}e^{ik_{y}y}\left(\begin{array}{c}
\cos\frac{\theta}{2}\phi_{N-1}\\
\sin\frac{\theta}{2}\phi_{N}
\end{array}\right)
\end{equation}
and for the - branch,
\begin{equation}
\psi_{N-}=\frac{1}{\sqrt{L_{y}}}e^{ik_{y}y}\left(\begin{array}{c}
-\sin\frac{\theta}{2}\phi_{N-1}\\
\cos\frac{\theta}{2}\phi_{N}
\end{array}\right)
\end{equation}
where $\theta$ is given by $\theta=\arctan\left[\frac{2\frac{\sqrt{2}}{R}\alpha \sqrt{N}}{g\mu_{B} B-\hbar w_{c}}\right]$
and $\phi$ is the normalized quantum harmonic oscillator wave function, i.e., Landau state, $N$ being the corresponding
Landau level index. According to these results the Rashba spin-orbit interaction mixes spin-down and
spin-up states of adjacent Landau
levels to give rise to two new energy branches of eigenstates of the Hamiltonian $H_{0}$.

To analyze magnetotransport in 2DES with RSOI we calculate the longitudinal
conductivity $\sigma_{xx}$ following the Boltzmann transport theory\cite{ridley,ando,askerov},
where $\sigma_{xx}$ is given by:
\begin{equation}
\sigma_{xx}=e^{2} \int_{0}^{\infty} dE \rho_{i}(E) [\Delta X(0)]^{2}W_{I}\left( -\frac{df(E)}{dE}  \right)
\end{equation}
being $E$ the energy, $\rho_{i}(E)$ the density of
initial states and $f(E)$ the electron distribution
function. $\Delta X(0)$ is the shift of the guiding center coordinate
for the eigenstates involved in the scattering event,\\
\begin{equation}
 \Delta X(0)=[X_{2}(0)-X_{1}(0)]\simeq 2R_{c}
 \end{equation}
$X_{2}(0)$ and $X_{1}(0)$ being the guiding center coordinates for final and initial states
respectively and $R_{c}$ the cyclotron radius. $W_{I}$ is the remote charged impurity scattering rate because
we consider that at very low temperatures ($T$) this is the most likely source of
scattering for electrons in high mobility 2DES.  According to
the Fermi's Golden Rule $W_{I}$ is given  by
\begin{equation}
W_{I}=\frac{2\pi}{\hbar}N_{I}|<\psi_{f\pm}|V_{s}|\psi_{i\pm}>|^{2}\delta(E_{f}-E_{i})
\end{equation}
where $N_{I}$ is the impurity density and  $E_{i}$ and $E_{f}$ are the energies of the initial and final states respectively.
$V_{s}$ is the scattering potential for charged impurities\cite{ando}.
The matrix element inside $W_{I}$ can be expressed as\cite{ridley,ando,askerov}:
\begin{equation}
|<\psi_{f\pm}|V_{s}|\psi_{i\pm}>|^{2}=\sum_{q}|V_{q}|^{2}|I_{if}|^{2}
\delta_{k^{'}_{y},k_{y}+q_{y}}
\end{equation}
where $V_{q}= \frac{e^{2}}{ \epsilon (q+q_{s})}$,
 $\epsilon$ the dielectric
constant and $q_{s}$ is the Thomas-Fermi screening
constant\cite{ando}. The integral  $I_{if}$ is given by:
\begin{equation}
I_{if}=\frac{1}{2}\int^{\infty}_{-\infty} [\pm\Phi_{f-1},\Phi_{f}]
\left(\begin{array}{clcr}
 e^{iq_{x}x}& 0\\
 0          & e^{iq_{x}x}
\end{array}\right)
\left[\begin{array}{c}
\pm\Phi_{i-1} \\
\Phi_{i}
\end{array}\right ]
 dx
\end{equation}
where we have considered that at low or moderate $B$ (used in experiments of
magnetoresistance oscillations) $\left[\frac{2\frac{\sqrt{2}}{R}\alpha \sqrt{N}}{g\mu_{B} B-\hbar w_{c}}\right]\rightarrow \infty$
and then $\theta\simeq \frac{\pi}{2}$

To calculate the density of states $\rho_{i}(E)$ of a 2DES with perpendicular $B$ and RSOI
we proceed starting off with the expression of the energy of the states, eq. (3) that can
be rewritten in a more compact way:
\begin{equation}
E_{N\pm}=\hbar w_{c}\left[N\pm \sqrt{\frac{1}{4}+\hbar 2 N \tilde{\alpha}^{2}}\right]
\end{equation}
where $\tilde{\alpha}^{2}=\alpha^{2}\frac{m^{*}}{\hbar^{4} w_{c}}$. To obtain the
new expression for $E_{N\pm}$ we have neglected the Zeeman term considering that at
the magnetic fields used in experiments and in simulations it is much
smaller than the Rashba term\cite{heida}. Expressing the density of states in terms of Dirac $\delta$-function
we can write:
\begin{eqnarray}
\rho_{i}(E)&=& \frac{eB}{h}\delta(E-E_{0})+\nonumber\\
           & &\frac{eB}{h}\sum_{N=1}^{\infty}[\delta(E-E_{N+})+\delta(E-E_{N-})]\nonumber\\
\end{eqnarray}
\\
where $E_{0}=\frac{\hbar w_{c}}{2}$.
To do the sum in the expression of $\rho_{i}$ we use the Poisson sum rules,
\begin{equation}
\sum_{n=1}^{\infty}f(n)=-\frac{1}{2}f(0)+\int_{0}^{\infty}f(x)dx+2\sum_{s=1}^{\infty}\int_{0}^{\infty}
\cos(2 \pi sx)f(x)dx
\end{equation}
and after some lengthy algebra we get to an expression that includes the state
broadening and reads\cite{amann,tarun}:
\begin{widetext}
\begin{equation}
\rho_{i}(E)=\frac{m^{*}}{\pi\hbar^{2}}\left \{ 1+\sum_{\pm}\left(1\pm \frac{\hbar \tilde{\alpha}^{2}}
{\sqrt{\frac{1}{4}+\frac{2E\tilde{\alpha}^{2}}{w_{c}}+\hbar^{2}\tilde{\alpha}^{4}}} \right)
\sum_{s=1}^{\infty}e^{\frac{-s\pi\Gamma}{\hbar w_{c}}}\cos\left[2\pi s\left(\frac{E}{\hbar w_{c}}+\hbar \tilde{\alpha}^{2}
\pm \sqrt{\frac{1}{4}+\frac{2E\tilde{\alpha}^{2}}{w_{c}}+\hbar^{2}\tilde{\alpha}^{4}} \right) \right]  \right \}
\end{equation}
$\Gamma$, being the sates width. This equation is essential in the present article because it
reveals the presence of two cosine terms that could interfere. On the other hand, it also
important to highlight that it is obtained
from an expression for the states energy that depends at the same time on the "Landau" level index, both
linearly and through a square root. With this expression of the states density we recover the
previous one obtained by Ch. Amann\cite{amann}  including the states broadening. This last condition
makes the expression much more useful to be used in theories explaining experimental results on 2DES with RSOI.  Considering that  only
electrons around the Fermi level participate in the magnetotransport and the usual electron
density used in these experiments\cite{ihn}, it turns out that
the $E$ term is much bigger than the Rashba term.
Therefore,
we can rewrite the expression of the density of states as:
\begin{equation}
\rho_{i}(E)=\frac{m^{*}}{\pi\hbar^{2}}\left \{ 1+
\sum_{s=1}^{\infty}e^{\frac{-s\pi\Gamma}{\hbar w_{c}}}\left[\cos2\pi s\left(\frac{E}{\hbar w_{c}}+
\sqrt{\frac{1}{4}+\frac{2E\tilde{\alpha}^{2}}{w_{c}}} \right)  + \cos 2\pi s\left(\frac{E}{\hbar w_{c}}-
\sqrt{\frac{1}{4}+\frac{2E\tilde{\alpha}^{2}}{w_{c}}} \right) \right] \right \}
\end{equation}
Finally and  after some algebra we can write an expression for $\sigma_{xx}$,
\begin{equation}
\sigma_{xx}=\frac{e^{2}m^{*}}{\pi\hbar^{2}} (\Delta X_{0})^{2} W_{I}\left \{ 1+
\sum_{s=1}^{\infty}e^{\frac{-s\pi\Gamma}{\hbar w_{c}}}\frac{X_{S}}{\sinh X_{S}}\left[\cos2\pi s\left(\frac{E_{F}}{\hbar w_{c}}+
\sqrt{\frac{1}{4}+\frac{2E_{F}\tilde{\alpha}^{2}}{w_{c}}} \right)  + \cos 2\pi s\left(\frac{E_{F}}{\hbar w_{c}}-
\sqrt{\frac{1}{4}+\frac{2E_{F}\tilde{\alpha}^{2}}{w_{c}}} \right) \right] \right \}
\end{equation}
\end{widetext}
where $E_{F}$ stands for the Fermi energy and $X_{S}=\frac{2\pi^{2}k_{B}T}{\hbar w_{c}}$, $k_{B}$ being
the Boltzmann constant.  To obtain
$R_{xx}$ we use the relation
$R_{xx}=\frac{\sigma_{xx}}{\sigma_{xx}^{2}+\sigma_{xy}^{2}}
\simeq\frac{\sigma_{xx}}{\sigma_{xy}^{2}}$, where
$\sigma_{xy}\simeq\frac{n_{i}e}{B}$ and
$\sigma_{xx}\ll\sigma_{xy}$, $n_{i}$ being the 2D
electron density.
The sum of cosine terms
in the expression of $\sigma_{xx}$ will give rise to an interference effect
that will become apparent as a beating pattern. Thus, the physical origin
of the beating pattern, that has been experimentally observed, can
be trace back to the slightly different energies of the two eigenstates branches.

The Hamiltonian $H_{0}$ is the same as the one of the surface states of nonideal
Dirac fermions in 3D topological insulators. The only difference is that for the latter
the quadratic term is small compared to linear term that it is the dominant when it comes to
topological insulators.
 In real samples the surface states of 3D topological insulators are no longer described by massles Dirac fermions.
 Experiments demonstrate important band bending and broken electron-hole symmetry with respect
 to the Dirac point in
 the band structure of real 3D topological insulators\cite{chen,ren}. Therefore the results presented above,
 especially the ones concerning density of states and states energy, could be of interest in
 the study of magnetotransport in real 3D topological insulators.

 If now we switch on radiation, first of all we have to add
 to the Hamiltonian $H_{0}$ a radiation term $H_{R}$ and then:
 $H_{0}=(H_{B}+H_{SO}+H_{R})$,
  where
\begin{equation}
H_{R}=-(x-X_{0})\varepsilon_{0}\cos wt -
 X_{0} \varepsilon_{0}\cos wt
\end{equation}
 $\varepsilon_{0}$ being the radiation
 electric field and $w$ the corresponding radiation frequency.
 $H_{0}$ can again be solved exactly \cite{ina2,ina3,kerner,park},
and the solution for the electronic
wave function is made up, as above, of two states branches.
The wave function for the + branch is,
\begin{equation}
\Psi_{N+}=\frac{1}{\sqrt{L_{y}}}e^{ik_{y}y}\left(\begin{array}{c}
\cos\frac{\theta}{2}\Psi_{N-1}(x,t)\\
\sin\frac{\theta}{2}\Psi_{N}(x,t)
\end{array}\right)
\end{equation}
and for the - branch,
\begin{equation}
\Psi_{N-}=\frac{1}{\sqrt{L_{y}}}e^{ik_{y}y}\left(\begin{array}{c}
-\sin\frac{\theta}{2}\Psi_{N-1}(x,t)\\
\cos\frac{\theta}{2}\Psi_{N}(x,t)
\end{array}\right)
\end{equation}
where,
\begin{eqnarray}
&&\Psi_{N}(x,t)=\Phi_{N}(x-X-x_{cl}(t),t)\nonumber  \\
&&\times  e^{ \left[i\frac{m^{*}}{\hbar}\frac{dx_{cl}(t)}{dt}[x-x_{cl}(t)]+
\frac{i}{\hbar}\int_{0}^{t} {\it L} dt'\right]}
\end{eqnarray}
as above, $\Phi_{n}$
is the solution for the Schr\"{o}dinger equation of the unforced
quantum harmonic oscillator where $x_{cl}(t)$ is the classical
solution of a forced harmonic oscillator\cite{ina2,kerner,park},
\begin{eqnarray}
x_{cl}(t)&=&\frac{e \varepsilon_{o}}{m^{*}\sqrt{(w_{c}^{2}-w^{2})^{2}+\gamma^{4}}}\cos ( wt-\beta)\nonumber\\
&=&A\cos ( wt-\beta)
\end{eqnarray}
$\gamma$ is a phenomenologically introduced damping factor
for the electronic interaction with acoustic phonons.
$\beta$ is the phase difference between the radiation-driven guiding center and
the driving radiation itself.
${L}$ with RSOI is now
given by,
\begin{equation}
L=\frac{1}{2}m^{*}\dot{x}_{cl}^{2}-\frac{1}{2}m^{*}w_{c}^{2}x_{cl}^{2}-\frac{\alpha}{\hbar}
\left[\sigma_{y}m^{*}\dot{x}_{cl}-\sigma_{x}eBx_{cl}\right]\nonumber\\
\\
\end{equation}
Apart from phase factors, the wave function for $H_{0}$ now is
the same as the standard harmonic oscillator where the center is
displaced by $x_{cl}(t)$.
\begin{figure}
\centering \epsfxsize=3.5in \epsfysize=6.0in
\epsffile{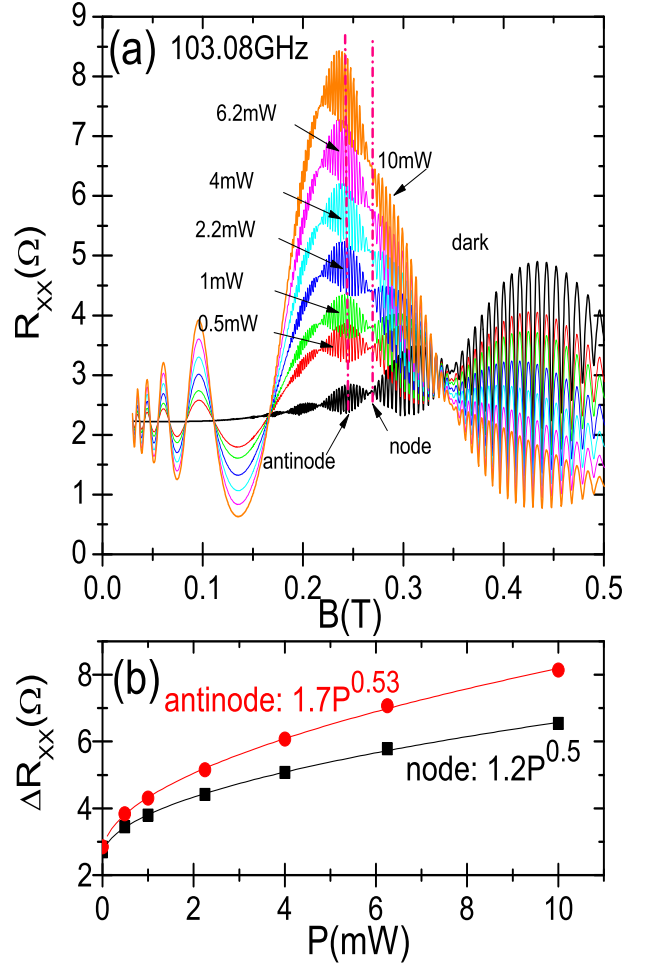}
\caption{Dependence on radiation power $P$ of the calculated magnetoresistivity under light
in 2DES with Rashba coupling. In panel (a) we exhibit $R_{xx}$ as a function
of $B$, for different radiation intensities starting from dark and for the same frequency
$f=103.08$ GHz. In panel (b) we exhibit  $\Delta R_{xx}= R_{xx}-R_{xx}(dark)$ versus $P$ for $B$
corresponding to dashed vertical lines on panel (a). One line corresponds to the $B$-position
of a node and the other of an antinode.  For both, node and antinode,
we obtain a square root (sublinear) dependence showing the corresponding
fits.(T=1K).}
\end{figure}
In the
presence of radiation, the electronic orbit center coordinates change and
are given according to our model by $X(t)=X(0)+x_{cl}(t)$.
This means that due to the radiation field all the electronic orbit
centers in the sample harmonically oscillate at the radiation frequency in the $x$ direction
through $x_{cl}$. Applying initial conditions, at $t=0$, $X(t)=X(0)$ and then
$\beta = \pi/2 $. As a result the expression for the time dependent
guiding center is now:
\begin{equation}
X(t)=X(0)+A \sin wt
\end{equation}
In the presence of charged impurities scattering and
radiation the average advanced distance by electrons is going to
be different than in the dark, $\Delta X(0)$ (see Fig. 1a). Now the positions
of the Landau states guiding centers are time-dependent according to
the last expression. If the scattering event begins at a certain time
$t$, the initial LS is given by, $X_{1}(t)=X_{1}(0)+A \sin wt$. After a time
$\tau$, that we call {\it flight time}, the electron "lands" in a final LS that is no longer $X_{2}$ as in the
dark scenario. All LS have been displaced in the same direction, the same distance  given by, $A \sin w\tau$.
Thus, due to the swinging nature of irradiated LS, its former
position is taken by another LS that we can call $X_{3}$ (see Fig. 1b). This
final LS is written as,  $X_{3}(t+\tau)=X_{3}(0)+A \sin w(t+\tau)$,
and the scattering-induced advanced distance by the electron reads,
\begin{eqnarray}
\Delta X(t)&=&X_{3}(t+\tau)-X_{1}(t)\nonumber\\
           &=&X_{3}(0)+A \sin w(t+\tau)-X_{1}(0)-A \sin wt\nonumber\\
\end{eqnarray}
In order to obtain the steady-state regime for the
advanced distance we time-average over a period of the radiation field,
\begin{eqnarray}
<\Delta X(t)>&=&<X_{3}(t+\tau)-X_{1}(t)>\nonumber\\
           &=&X_{3}(0)+<A \sin w(t+\tau)>-\nonumber\\
           &&X_{1}(0)-<A \sin wt>\nonumber\\
           &=&X_{3}(0)-X_{1}(0)
\end{eqnarray}
where obviously we have taken into account that
$<A \sin w(t+\tau)>=0$ and $<A \sin wt>=0$. Here, the angular brackets
describe time-average over a period of the time-dependent field.
Next, we have to relate $<\Delta X(t)>$ with the advanced
distance in the dark $\Delta X(0)$. This is straightforward considering
first, that during the time $\tau$ all LS have been displaced in phase
the same distance $A \sin w\tau$, and secondly, that the initially
position occupied by $X_{2}(t)$ is now occupied after $\tau$ by $X_{3}(t+\tau)$.
In other words,  $X_{3}$ is the only orbit  always located at a distance $A \sin w\tau$
from $X_{2}(t)$.
Thus, we necessarily conclude that the respective guiding centers of
both LS $X_{3}(0)$ and $X_{2}(0)$ are separated by the distance, $A \sin w\tau$,
i.e.,  $X_{3}(0)=X_{2}(0)-A \sin w\tau$ (see Fig. 1c).
Substituting this result in the above expression we obtain,
\begin{eqnarray}
<\Delta X(t)>=\Delta X(\tau)&=&X_{2}(0)-A \sin w\tau-X_{1}(0)\nonumber\\
&=&\Delta X(0)-A \sin w\tau
\end{eqnarray}
According to our model, this expression is responsible of RIRO including the
maxima and minima positions. In order to obtain an expression and a
physical meaning for $\tau$,  we can compare the condition
fulfilled by the minima positions obtained from the theoretical  expression to
the one obtained in experiments\cite{mani1}. These minima positions
represent one of the main traits describing RIRO and were first found by
Mani et al\cite{mani1} being given by:
\begin{equation}
\frac{w}{w_{c}}=\frac{5}{4},\frac{9}{4},\frac{13}{4}....=\left(\frac{1}{4}+ n \right)
\end{equation}
\\
where $n=1,2,3....$.
According to theory, i.e., expression (26), the minima positions
are obtained when,
\begin{equation}
w\tau = \frac{\pi}{2} + 2\pi n \Rightarrow w=\frac{2\pi}{\tau}\left(\frac{1}{4}+ n \right)
\end{equation}
Then, comparing both expressions we readily obtain that the flight time is,
\begin{equation}
\tau=\frac{2\pi}{w_{c}}
\end{equation}
In other words, $\tau$ equals the cyclotron period $T_{c}$.
This important result admits a semiclassical approach in the sense that during
the time it takes the electron to "fly"  from one LS to another due to scattering, the electrons
in their orbits perform one whole loop.

Finally, the advanced distance due to scattering in the presence
of radiation reads,
\begin{equation}
\Delta X(\tau)= \Delta X(0)-A\sin \left(2\pi\frac{w}{w_{c}}\right)
\end{equation}

Applying these last results   to a Boltzmann transport model, similarly
as the first part of this section,
we can get to an expression for the longitudinal conductivity of the magnetotransport
of a high mobility 2DES with strong RSOI in the presence of radiation. In this expression three harmonic terms turn up, two cosine
terms depending on the Fermi energy and $\alpha$, that interfere to give rise to the beating pattern profile in the
magnetoresistance.  And one sine term depending on radiation parameters, frequency and power. We expect the latter
to interfere on the beating pattern profile.
\begin{widetext}
\begin{equation}
\sigma_{xx}\propto  \left[\Delta X^{0}-A\sin \left(2\pi\frac{w}{w_{c}}\right)\right]^{2} \left \{ 1+
e^{\frac{-\pi\Gamma}{\hbar w_{c}}}\frac{X_{S}}{\sinh X_{S}}\left[\cos2\pi \left(\frac{E_{F}}{\hbar w_{c}}+
\sqrt{\frac{1}{4}+\frac{2E_{F}\tilde{\alpha}^{2}}{w_{c}}} \right)  + \cos 2\pi \left(\frac{E_{F}}{\hbar w_{c}}-
\sqrt{\frac{1}{4}+\frac{2E_{F}\tilde{\alpha}^{2}}{w_{c}}} \right) \right] \right \}
\nonumber\\
\\
\end{equation}
\end{widetext}

where we have considered only the term $s=1$, the most important, in the sum.
We consider that the above results can be of application and can predict the
behavior of  magnetotransport in
3D topological insulators subjected to radiation;  once these systems reach enough mobility to make
patent the rise of RIRO. This could happen at the same time that, without radiation,
the $R_{xx}$ beating patter begins to be visible in magnetotransport experiments in 3D topological
insulators.
\\
\begin{figure}
\centering \epsfxsize=3.5in \epsfysize=6.0in
\epsffile{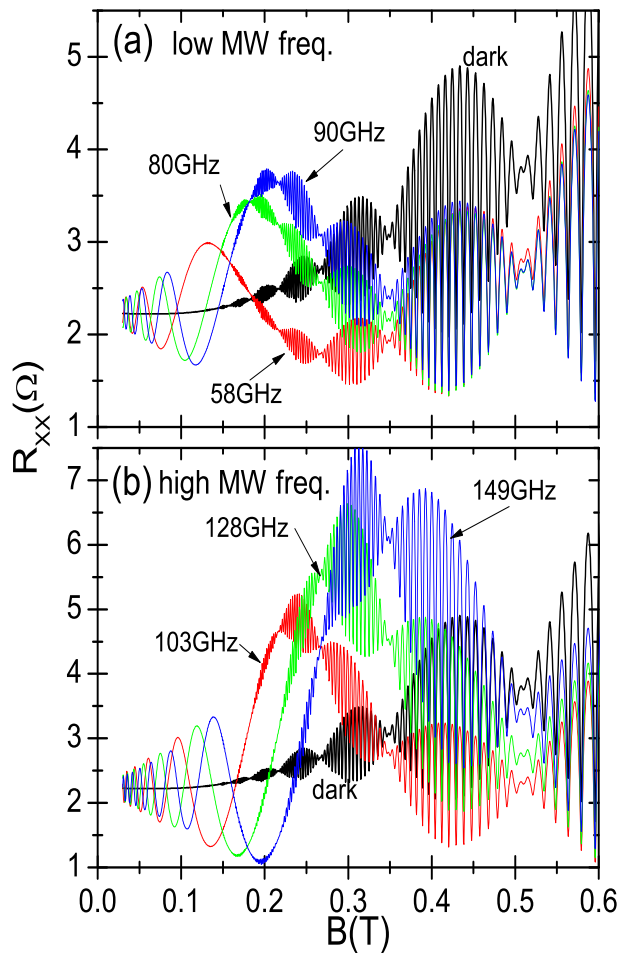}
\caption{Dependence on radiation frequency, $f$ of irradiated $R_{xx}$ vs $B$ for
2DES with Rashba coupling. In panel (a) we exhibit the low frequency scenario and in panel (b)
the high frequency, obtaining similar results for both. Nodes $B$-position is
immune to frequency and antinodes shape depends on frequency because the former
depends on RIRO's positions that, in turn, deeply depend on $f$. (T=1K).}
\end{figure}
\section{Results}
All calculated results presented in this article are based on the next list of
parameteres regarding experiments in InAs quantum wells\cite{ihn,nitta,shoja,heida,luo,luo2}: Rashba parameter
$\alpha=0.6\times 10^{-11}eV\cdot m$, electron density $n_{i}=2.0\times10^{16}m^{-2}$,
electron effective mass $m^{*}=0.045m_{e}$ where $m_{e}$ is the
electron rest mass and temperature, $T=1K$.\\
In Fig.(1) we present calculated  $R_{xx}$ vs $B$ for dark and irradiated
scenarios in a high-mobility 2DES with strong RSOI.
 For the latter, the radiation frequency $f= 149$ GHz.
For the dark curve we obtain a very clear beating pattern profile made up of
a system of nodes and antinodes. The radiation curve exhibits a similar beating
pattern but this time dramatically deformed and modulated by the rise
of the system of peaks and valleys of RIRO. In the new beating pattern
the node $B$-positions are not affected by the presence of radiation but yet the
different antinodes are, according to their $B$-position. This peculiar profile  in $R_{xx}$
shows up as result of the interference effect between the sine and
cosine terms that is reflected in
equation (23).

In Fig.(2) we present  the dependence of calculated $R_{xx}$ on $P$ for 2DES with
important Rashba coupling under radiation. In panel (a)
we exhibit calculated $R_{xx}$ vs $B$ for a radiation frequency $f=103.08$GHz, different  radiation intensities from dark
to $10$ mW: $0.5$,  $1$, $2.2$, $4$, $6.2$ and $10$ mW and $T=1$K.
We easily observe, as expected, that RIRO increase their amplitudes as $P$ increases from dark.
At the same time the deformation of antinodes gets stronger too, keeping constant the
$B$-position of the nodes. In panel (b) we exhibit, again for $f=103.08$GHz,  $\Delta R_{xx}= R_{xx}-R_{xx}(dark)$ versus $P$ for $B$
corresponding to dashed vertical lines on panel (a). One line corresponds to the $B$-position
of a node and the other of an antinode. We want to check out if the presence
of Rashba coupling affects the  previously obtained sublinear power law for
the dependence of RIRO on $P$. In this way we obtain for both, according to
the calculated fits (see Fig. (2.b)), an approximately square root
dependence on $P$,  concluding that Rashba coupling
does not affect the sublinear law.
We can theoretically explain these results according to our model.
In the expression of $\sigma_{xx}$ and then in $R_{xx}$, $P$ only shows up in the numerator of the amplitude $A$
 as $\sqrt{P}\propto \varepsilon_{0}$,
but not in the phase of the sine function.
Thus, on the one hand, $P$ does not affect the phase of RIRO that remains constant as $P$ changes,
and on the other hand $ R_{xx}\propto \sqrt{P}$, giving rise to the sublinear (square root)
power law for the dependence of $R_{xx}$ on $P$. Finally, in the
phase of cosine terms there is no radiation parameters concluding
that radiation will not affect the $B$-positions of
nodes and antinodes.
\begin{figure}
\centering \epsfxsize=3.5in \epsfysize=5.5in
\epsffile{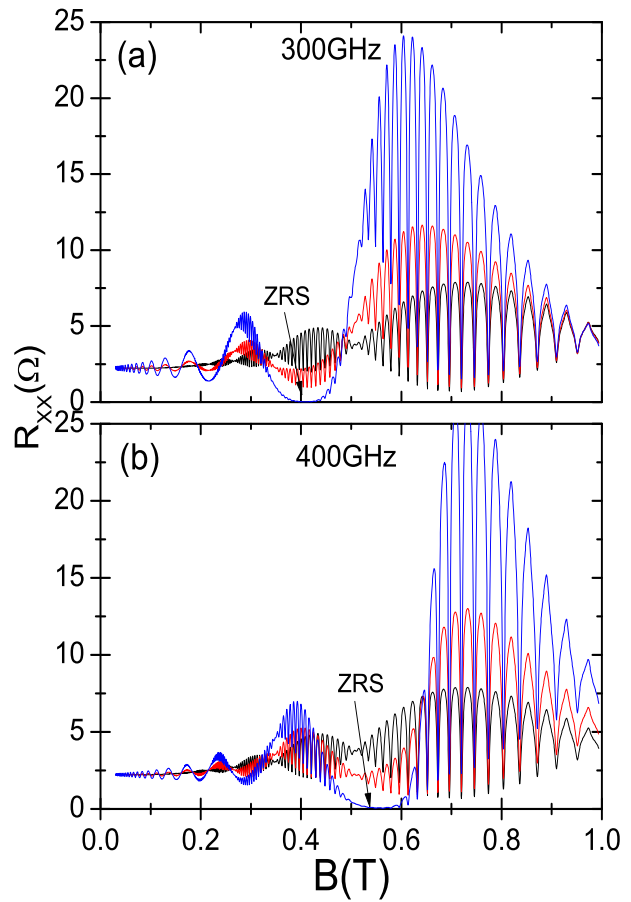}
\caption{ Terahertz irradiated $R_{xx}$ versus $B$ for 2DES with Rashba interaction
 for different temperatures with constant power
excitation for $f=300$GHz in panel (a) and $f=400$GHz in panel (b).
We obtain a ZRS region in each panel as indicated by arrows.
In panel (a) an antinode is wiped out immersed in the
ZRS region as the radiation intensity increases. The same in
panel (b) but now for a node. (T=1K).}
\end{figure}

In Fig.(3) we present the dependence on radiation frequency of
irradiated $R_{xx}$ for 2DES with Rashba interaction.
In panel (a) we show the low frequency case and in panel (b) the high frequency, obtaining
similar results for both. Thus, the nodes $B$-position turns out
to be immune to radiation frequency keeping the same ones as in the dark situation.
The deformation of the antinodes changes with the frequency. The reason
is that the deformation or modulation depends on the RIRO position and
the latter does change with radiation frequency. As a result, the same
initial antinode in the dark will deform differently according to $f$. We also observe
that the strongest deformation corresponds to the RIRO's peaks irrespective
of radiation frequency. The immunity of nodes with $f$ can be readily explained
as in the previous figure,
according to Eq. (23). In this equation the nodes position depends only
on the cosine terms where the Rashba term $\alpha$ shows up in the corresponding phases.
In these phases $f$ does not show up  and then its variation
will not affect the positions of either the nodes or the antinodes.

In Fig.(4) we present the obtained results for irradiated $R_{xx}$ vs $B$
for 2DES with Rashba coupling in the terahertz regime
showing two frequencies: $300$ GHz in panel (a) and $400$ GHz in
panel (b). For both panels we exhibit the dark case and two radiation
curves. For the latter, one is obtained at low radiation
intensity and the other at high. Apart form RIRO, due to radiation, and the beating
patter, due to Rashba,  we have obtained zero resistance states for
$B\simeq 0.4$T in the upper panel and for $B\simeq 0.55$T in the lower
panel. In the former case ZRS are  obtained increasing $P$ from an antinode in the
dark scenario. This antinode ends up totally wiped out as ZRS
rise up. Similar situation is presented in the lower panel but
this time ZRS is obtained from a node. Similarly as before,
the node disappears immersed in the ZRS region.

\section{Conclusions}
In summary we have  presented a theoretical analysis on the effect of radiation
on  mangetotransport  of 2DES with
strong Rashba spint-orbit coupling. We  have studied the interaction between
the radiation-induced
resistance oscillations and the typical beating pattern showing up
in the mangetoresistance of 2DES with Rashba interaction.
We have deduced an exact
solution for the electron wave function corresponding to a total Hamiltonian with
 Rashba coupling  and radiation terms.
 We have developed
 a perturbation treatment for elastic scattering due
to charged impurities based on the Boltzmann theory to finally obtain an
expression for the
magnetoresistance of the system.  In the presence of radiation we
have obtained that the typical
 beating pattern of a 2DES with Rashba interaction is strongly modified following the profile of radiation-induced
magnetoresistance oscillations. We have studied also
the dependence on  intensity and frequency of
radiation, including the teraherzt regime. For this regime
we have also obtained ZRS beginning from the dark in the case of
a node and in the case of antinode. We think that the results presented
in this paper,  could be of
interest for magnetotransport of nonideal Dirac fermions in 3D topological insulators
considering that both systems share similar Hamiltonian. The only difference
is that for 3D topological insulators the quadratic term is smaller than the linear that in this case
is dominant. When 3D topological insulators reach enough electron mobility, all the effects
related above, from the beating pattern to its modulation under the
presence of radiation will be clearly apparent.
\\

\section{Acknowledgments}
We acknowledge V. Tribaldos and J.M. Reynolds for useful discussions.
This work is supported by the MINECO (Spain) under grant
MAT2014-58241-P  and ITN Grant 234970 (EU).
GRUPO DE MATEMATICAS APLICADAS A LA MATERIA CONDENSADA, (UC3M),
Unidad Asociada al CSIC.

\end{document}